# Asymptotic analysis of the dispersion relation of an incompressible elastic layer of uniform thickness.

## W. Hussain*

Department of Mathematics, School of Arts and Sciences, Lahore University of Management Sciences (LUMS), Opposite Sector'U', D.H.A., Lahore Cantt. 54792, Pakistan.

* *E-mail address*: wasiq@lums.edu.pk (W.Hussain)
Tel: 92-42-5722670-79(10 lines); fax: 92-42-5722591

## Abstract

This paper is concerned with an *asymptotic analysis* of the dispersion relation for wave propagation in an elastic layer of uniform thickness. The layer is subject to an underlying simple shear deformation accompanied by an arbitrary uniform hydrostatic stress. In respect of a general form of incompressible, isotropic elastic strain-energy function, the dispersion equation for infinitesimal waves is obtained *asymptotically* for incremental traction boundary conditions on the faces of the layer. Two sets of wave speeds as a function of wave number, layer thickness and material parameter are obtained and the numerical results are illustrated for two different classes of strain-energy functions. For a particular class of strain-energy functions, it is shown that wave speeds don't depend upon the amount of shear and the material parameters. For materials not in this special class velocities do depend upon the shear deformation.

## 1. Introduction

The study of infinitesimal waves propagating in finitely deformed elastic material was initiated in a series of papers by Biot, summarized in his monograph [1], and by Hayes and Rivlin [2], who studied surface waves in a half-space subject to pure homogeneous strain. Dowaikh and Ogden [3,4] for incompressible and compressible isotropic elastic materials respectively, have investigated surface waves and deformations of a half-space, again for a finite deformation corresponding to pure homogeneous strain. In [11,12] the reflection of infinitesimal plane waves from the plane boundary of a half-space subject to pure homogeneous strain is worked out. Discussion of Flexural waves in incompressible pre-stressed elastic composites and small amplitude vibrations of pre-stressed laminates is done by Rogerson and Sandiford in [13] and [14] respectively.

In each of the papers cited above, the considered finite deformation is a pure homogeneous strain so that the orientation of the principal axes of strain is fixed. The effect of principal axis orientation of waves and deformations in a plate or half-space has been exemplified in the case of simple shear in Refs. [5,6]. More recently, Hussain and Ogden [7] did the problem of reflection and transmission of plane waves, in which the two half-spaces consist of the same incompressible isotropic elastic material but are subject to equal and opposite simple shears (see also the related paper [8]).

In [6] Connor and Ogden examined the propagation of dispersive waves in a layer, of finite thickness, which is subject to a plane deformation consisting of simple shear parallel to the faces of the layer. As a consequence dispersion equation was derived and analyzed *non-asymptotically* for incremental boundary conditions.

In this paper the analysis in [6] is extended and the *dispersion relation* is *analyzed asymptotically* and the numerical results for the large wave velocities and low wave numbers are obtained.

The required notations and equations are summarized in Section 2. In Section 3 the brief derivation of the dispersion equation from [6] is given relevant to the incremental traction boundary conditions.

In Section 4 the *asymptotic expansion* of the dispersion equation is derived, and by using the asymptotic solutions of the propagation condition, two sets of velocities for different modes as a function of wave number, layer thickness and material parameter are obtained.

Graphical results are provided in Section 5 in order to illustrate the dependence of the wave velocities on the wave number and layer thickness for the two categories of strain energy functions. The features of the results are described for the two classes of strain-energy functions in Sections 5.1 and 5.2. Unlike [6], for the special class of strain-energy functions, the wave velocities are independent of the simple shear deformation. For materials not in this special class, on the other hand, it is shown that the wave velocities do depend upon the amount of shear. In both classes of strain-energy functions, the wave velocities are independent of the hydrostatic stress, which is different from the previous work [6].

## 2. Basic equations

Let **X** and **x**, respectively, be the position vectors of a typical material particle in the undeformed (reference) and deformed configurations $B_0$ and $B$ of the material. We write the deformation, **c** say, in the form

$$\mathbf{x} = \mathbf{c}(\mathbf{X}), \quad \mathbf{X} \in B_0. \tag{2.1}$$

The deformation gradient **A** is defined by

$$\mathbf{A} = \text{Grad}\,\mathbf{c}, \tag{2.2}$$

where the gradient operator Grad is with respect to $B_0$.

With reference to Cartesian basis vectors $\{\mathbf{E}_\alpha\}$ and $\{\mathbf{e}_i\}$ associated with $B_0$ and $B$ respectively, we may write

$$\mathbf{A} = A_{i\alpha}\mathbf{e}_i \otimes \mathbf{E}_\alpha, \quad A_{i\alpha} = \frac{\partial x_i}{\partial X_\alpha}, \tag{2.3}$$

with summation over $i$ and $\alpha$ from 1 to 3 implied in the first equation in Eq. (2.3), where $A_{i\alpha}$ are the Cartesian components of $\mathbf{A}$. The usual requirement

$$\det \mathbf{A} > 0 \tag{2.4}$$

is assumed to hold.

From the polar decomposition theorem we have

$$\mathbf{A} = \mathbf{V}\mathbf{R}, \tag{2.5}$$

where $\mathbf{R}$ is a proper orthogonal tensor and $\mathbf{V}$ (*the left stretch tensor*) is symmetric and positive definite.

The spectral decomposition of $\mathbf{V}$ has the form

$$\mathbf{V} = \sum_{i=1}^{3} \lambda_i \mathbf{v}^{(i)} \otimes \mathbf{v}^{(i)}, \tag{2.6}$$

where $\lambda_i (>0) (i=1,2,3)$ are the principal stretches and $\mathbf{v}^{(i)}$ $(i=1,2,3)$ the (unit) Eulerian principal axes.

It follows from Eqs. (2.5) and (2.6) that

$$\mathbf{A}\mathbf{A}^{\mathrm{T}} = \mathbf{V}^2 = \sum_{i=1}^{3} \lambda_i^2 \mathbf{v}^{(i)} \otimes \mathbf{v}^{(i)}, \tag{2.7}$$

where $\mathbf{A}^{\mathrm{T}}$ denotes the transpose.

For an isochoric deformation we have

$$\det \mathbf{A} \equiv l_1 l_2 l_3 = 1. \tag{2.8}$$

For an isotropic elastic material with strain-energy function $W$ per unit volume, $W$ depends symmetrically on $l_1, l_2, l_3$. We take the material to be incompressible, so that Eq. (2.8) holds identically. The associated principal Cauchy stresses $s_1, s_2, s_3$ are then given by

$$s_i = l_i \frac{\partial W}{\partial l_i} - p, \quad i \in \{1,2,3\}, \tag{2.9}$$

where $p$ is the arbitrary hydrostatic stress associated with the incompressibility constraint.

Let $\mathbf{S}$ denote the nominal stress tensor. Then, in the absence of body forces, the equilibrium of the body is governed by the equation

$$\text{Div}\,\mathbf{S} = \mathbf{0}, \tag{2.10}$$

where Div is the divergence operator in $B_0$.

For a homogeneous incompressible elastic material we have

$$\mathbf{S} = \frac{\partial W}{\partial \mathbf{A}} - p\mathbf{A}^{-1}. \tag{2.11}$$

For convenience we take $\mathbf{e}_1 = \mathbf{E}_1$, $\mathbf{e}_2 = \mathbf{E}_2$, $\mathbf{e}_3 = \mathbf{E}_3$. If the deformation is confined to (1,2)-plane, we have $\mathbf{v}^{(3)} = \mathbf{e}^{(3)}$, and we may set $\lambda_3 = 1$, so that Eq. (2.8) reduces to

$$l_1 l_2 = 1. \tag{2.12}$$

Let $f$ denote the orientation of the Eulerain axes ($\mathbf{v}^{(1)}, \mathbf{v}^{(2)}$) relative to the Cartesian axes ($\mathbf{e}_1, \mathbf{e}_2$) so that the Eulerian coordinates ($x_1', x_2'$) are related to the coordinates ($x_1, x_2$) associated with ($\mathbf{e}_1, \mathbf{e}_2$) by

$$x_1' = x_1 \cos f + x_2 \sin f, \qquad x_2' = -x_1 \sin f + x_2 \cos f. \tag{2.13}$$

For simple shear deformation, we may write

$$\lambda_1 = \lambda, \quad \lambda_2 = \lambda^{-1}, \quad \varepsilon = \lambda - \lambda^{-1}, \qquad (2.14)$$

where $\varepsilon$ is the amount of shear and the angle $\phi$ in Eq. (2.13) is given by

$$\tan 2\phi = \frac{2}{\varepsilon}, \qquad (2.15)$$

with $0 < \phi \leq \frac{\pi}{4}$ for $\varepsilon \geq 0$ and $\frac{\pi}{4} \leq \phi < \frac{\pi}{2}$ for $\varepsilon \leq 0$. From Eq. (2.15) we deduce that

$$\cos\phi = \frac{\lambda}{\sqrt{\lambda^2+1}}, \quad \sin\phi = \frac{1}{\sqrt{\lambda^2+1}}. \qquad (2.16)$$

Superimposed on the finite deformation just described we consider a small time-dependent displacement **u** with components referred to $\{\mathbf{e}_i\}$ such that

$$u_1 = u_1(x_1, x_2, t), \quad u_2 = u_2(x_1, x_2, t), \quad u_3 = 0, \qquad (2.17)$$

where $t$ is time.

Let $\Sigma$ denote the increment in nominal stress associated with this motion relative to finitely deformed configuration. Then, the (linearized) stress-deformation relation, in components referred to the basis $\{\mathbf{e}_i\}$, has the form

$$\Sigma_{ji} = A_{0jilk} u_{k,l} + p u_{j,i} - \pi \delta_{ij}, \qquad (2.18)$$

where $A_{0jilk}$ are the components of the tensor of (first-order) instantaneous elastic moduli referred to $B$, $\pi$ is the increment in $p$, $\delta_{ij}$ is the Kronecker delta and $,i$ signifies differentiation with respect to $x_i$. See, for example, Ref.[9] for details.

The corresponding incremental form of the incompressibility condition is

$$u_{i,i} = 0. \qquad (2.19)$$

The equation governing the incremental motion is

$$\Sigma_{ji,j} = \rho \ddot{u}_i, \qquad (2.20)$$

where $\rho$ is the mass density of the material and a superposed dot indicates the material time derivative. On substitution of Eq. (2.18) in Eq. (2.20) and using Eq. (2.19) we obtain

$$A_{0jilk} u_{k,jl} - \pi_{,i} = \rho \ddot{u}_i. \qquad (2.21)$$

When specialized to the plane motion considered here Eq. (2.21) applies for $i = 1,2$ and summation run over indices in $\{1,2\}$. Elimination of $\pi$ between these two equations by cross differentiation and subtraction yields

$$A_{0j1lk}u_{k,2jl} - A_{0j2lk}u_{k,1jl} = \rho(\ddot{u}_{1,2} - \ddot{u}_{2,1}). \tag{2.22}$$

It is convenient to re-cast Eq. (2.22) in terms of components relative to the Eulerian axes with coordinates $(x_1', x_2')$. Let $u_i'$ and $A'_{0jilk}$ be the components in this case, with $u_i'$ regarded as functions of $(x_1', x_2', t)$. From the incompressibility condition Eq. (2.19) in terms of these components we deduce the existence of a function $\psi' = \psi'(x_1', x_2', t)$ such that

$$u_1' = \psi'_{,2}, \qquad u_2' = -\psi'_{,1}, \tag{2.23}$$

wherein the indicated differentiations are with respect to $(x_1', x_2')$.

When referred to the Eulerian axes Eq. (2.22), after substituting from Eq. (2.23), yields an equation for $\psi'$, namely

$$\alpha\psi'_{,1111} + 2\beta\psi'_{,1122} + \gamma\psi'_{,2222} = \rho(\ddot{\psi}'_{,11} + \ddot{\psi}'_{,22}), \tag{2.24}$$

as in Refs.[5,6], where the material parameters $\alpha, \beta, \gamma$ are defined by

$$\alpha = A'_{01212}, \quad \gamma = A'_{02121}, \quad 2\beta = A'_{01111} + A'_{02222} - 2A'_{01122} - 2A'_{01221}. \tag{2.25}$$

The associated strong-ellipticity condition consists of the inequalities

$$\alpha > 0, \quad \gamma > 0, \quad \beta > -\sqrt{\alpha\gamma}. \tag{2.26}$$

Here, the components $A'_{0jilk}$ are constants since the underlying deformation is homogeneous For future reference we note that the components $A_{0jilk}$ are related to $A'_{0jilk}$ through

$$A_{0ijkl} = l_{pi}l_{qj}l_{rk}l_{sl} A'_{0pqrs} \tag{2.27}$$

where

$$(l_{ij}) = \begin{pmatrix} \cos\phi & \sin\phi \\ -\sin\phi & \cos\phi \end{pmatrix} \tag{2.28}$$

and we recall Eq. (2.16).

Eq. (2.24) may also be expressed in terms of derivatives with respect to $(x_1, x_2)$ by use of Eq. (2.13) and the notational change $\mathbf{y}(x_1, x_2, t) \equiv \mathbf{y}'(x_1', x_2', t)$. Explicitly, we then have

$$(\mathbf{a}\cos^4\mathbf{f} + 2\mathbf{b}\sin^2\mathbf{f}\cos^2\mathbf{j} + \mathbf{g}\sin^4\mathbf{f})\mathbf{y}_{,1111} + 2\sin 2\mathbf{f}(\mathbf{a}\cos^2\mathbf{f} - \mathbf{b}\cos 2\mathbf{j} - \mathbf{g}\sin^2\mathbf{f})\mathbf{y}_{,1112} +$$

$$[2\mathbf{b} + 6(\mathbf{a} + \mathbf{g} - 2\mathbf{b})\sin^2\mathbf{f}\cos^2\mathbf{j}]\mathbf{y}_{,1122} + 2\sin 2\mathbf{f}(\mathbf{a}\sin^2\mathbf{f} + \mathbf{b}\cos 2\mathbf{j} - \mathbf{g}\cos^2\mathbf{f})\mathbf{y}_{,1222} +$$

$$(\mathbf{a}\sin^4\mathbf{f} + 2\mathbf{b}\sin^2\mathbf{f}\cos^2\mathbf{j} + \mathbf{g}\cos^4\mathbf{f})\mathbf{y}_{,2222} = \mathbf{r}(\ddot{\mathbf{y}}_{,11} + \ddot{\mathbf{y}}_{,22}). \tag{2.29}$$

### 3. Dispersion equation

*In this section a brief review of the dispersion equation derived in Ref.[6] is given.*
Consider time-harmonic waves propagating in the $x_1$ direction, with speed $c$, of the form

$$\mathbf{y} = A\exp i[ks x_2 + \mathbf{w}t - kx_1], \tag{3.1}$$

where $A$ is a constant, $\mathbf{w}$ the frequency, $k = \dfrac{\mathbf{w}}{c}$ and $s$, which is in general complex.

On substituting Eq. (3.1) in Eq. (2.29), it follows that $s$ satisfies the equation

$$(\mathbf{a}\sin^4\mathbf{f} + 2\mathbf{b}\sin^2\mathbf{f}\cos^2\mathbf{j} + \mathbf{g}\cos^4\mathbf{f})s^4 - 2\sin 2\mathbf{f}(\mathbf{a}\sin^2\mathbf{f} + \mathbf{b}\cos 2\mathbf{j} - \mathbf{g}\cos^2\mathbf{f})s^3 +$$

$$[2\mathbf{b} + 6(\mathbf{a} + \mathbf{g} - 2\mathbf{b})\sin^2\mathbf{f}\cos^2\mathbf{j} - \mathbf{r}c^2]s^2 - 2\sin 2\mathbf{f}(\mathbf{a}\cos^2\mathbf{f} - \mathbf{b}\cos 2\mathbf{j} - \mathbf{g}\sin^2\mathbf{f})s$$

$$+ \mathbf{a}\cos^4\mathbf{f} + 2\mathbf{b}\sin^2\mathbf{f}\cos^2\mathbf{j} + \mathbf{g}\sin^4\mathbf{f} - \mathbf{r}c^2 = 0. \tag{3.2}$$

Alternatively, since $\cot \mathbf{f} = \mathbf{l}$ follows from Eq. in (2.16), equation Eq. (3.2) may be written more compactly as

$$s^4 - 2\mathbf{e}s^3 + [4\mathbf{d} + 2 + \mathbf{e}^2 - (1+\mathbf{d})\mathbf{z}]s^2 - 2(1+2\mathbf{d})\mathbf{e}s + 1 + (1+\mathbf{d})\mathbf{e}^2 - (1+\mathbf{d})\mathbf{z} = 0, \tag{3.3}$$

where the *material* parameter $\mathbf{d}$ is defined by

$$\mathbf{d} = \frac{\mathbf{a} + \mathbf{g} - 2\mathbf{b}}{2(\mathbf{b} + \sqrt{\mathbf{ag}})}, \tag{3.4}$$

and $z$ is a non-dimensional squared wave speed defined by

$$z = \frac{rc^2}{\sqrt{ag}}. \tag{3.5}$$

We note that

$$d + 1 \equiv \frac{(\sqrt{a} + \sqrt{g})^2}{2(b + \sqrt{ag})} > 0 \tag{3.6}$$

is a consequence of Eq. (2.26).

Let the roots of Eq. (3.3) be denoted by $s_1, s_2, \bar{s}_1, \bar{s}_2$, where the over bar denotes complex conjugate. The appropriate form of $y$ is now

$$y = (A_1 e^{iks_1 x_2} + A_2 e^{iks_2 x_2} + B_1 e^{ik\bar{s}_1 x_2} + B_2 e^{ik\bar{s}_2 x_2}) e^{i(wt - kx_1)}, \tag{3.7}$$

where $A_1, A_2, B_1, B_2$ are constants.

Introducing *incremental traction boundary conditions* as

$$\Sigma_{21} = \Sigma_{22} = 0, \tag{3.8}$$

where $\Sigma_{21}$ and $\Sigma_{22}$ are the components of $\Sigma$ with respect to the ($x_1, x_2$) coordinate axes and can be obtained from Eq. (2.18). See Ref.[8] for details.

Consider an elastic layer of finite thickness (say $h$), and by using Eq. (3.7) and Eq. (3.8) on $x_2 = 0, -h$, we obtain the following *dispersion equation*, which we express in the form

$$(f_1 \bar{g}_1 - \bar{f}_1 g_1)(f_2 \bar{g}_2 - \bar{f}_2 g_2)(\bar{a}_1 a_2 + a_1 \bar{a}_2 - 1 - a_1 \bar{a}_1 a_2 \bar{a}_2) + \\ (f_1 \bar{g}_2 - \bar{f}_2 g_1)(\bar{f}_1 g_2 - f_2 \bar{g}_1)(\bar{a}_1 a_1 + a_2 \bar{a}_2 - 1 - a_1 \bar{a}_1 a_2 \bar{a}_2) = 0. \tag{3.9}$$

where $a_1 = e^{-ikhs_1}$, $a_2 = e^{-ikhs_2}$, $\bar{a}_1 = e^{ikh\bar{s}_1}$, $\bar{a}_2 = e^{ikh\bar{s}_2}$,

$$f_i = f(s_i), \quad \bar{f}_i = f(\bar{s}_i), \quad \bar{g}_i = g(\bar{s}_i), \quad g_i = g(s_i) \quad i \in \{1,2\}, \tag{3.10}$$

and the values of $f$ and $g$, by using Eq. (3.6), in case of simple shear are given by

$$f(s) = (1 + d)^{-1}(s - l)(s + l^{-1})^2 + 3s - l + 2l^{-1} + \bar{p}(s - l) - zs, \tag{3.11}$$

$$g(s) = (1+\boldsymbol{d})^{-1}(s-\boldsymbol{1})^2(s+\boldsymbol{1}^{-1}) + 3s - 2\boldsymbol{1} + \boldsymbol{1}^{-1} + \overline{p}(s+\boldsymbol{1}^{-1}) - \boldsymbol{z}s, \qquad (3.12)$$

where

$$\overline{p} = \frac{(\boldsymbol{g}-\boldsymbol{s}_2)}{\sqrt{a\boldsymbol{g}}}. \qquad (3.13)$$

Reader is advised to see [6] for the detailed derivations of Eqs. (3.9), (3.11) and (3.12).

The dependence on the strain-energy function arises through the parameter $\boldsymbol{d}$ in Eq. (3.3) and through $\boldsymbol{d}$ and $\overline{p}$ in Eqs. (3.11) and (3.12).

## 4. Asymptotic analysis of the dispersion equation

In [6] the dispersion equation Eq. (3.9) is analyzed *non-asymptotically*. In this section we shall do the asymptotic analysis of Eq. (3.9).

Using $s = -iq,$ \hfill (4.1)

and rewriting Eq. (3.3) as

$$q^4 - 2\boldsymbol{e}iq^3 + (a\boldsymbol{z} - b)q^2 + 2i(1+2\boldsymbol{d})\boldsymbol{e}q + d - a\boldsymbol{z} = 0, \qquad (4.2)$$

where

$$a = 1+\boldsymbol{d}, \quad b = 4\boldsymbol{d} + 2 + \boldsymbol{e}^2, \quad \text{and} \quad d = 1+(1+\boldsymbol{d})\boldsymbol{e}^2. \qquad (4.3)$$

Introducing the perturbation technique

$$q = vq^{(1)} + q^{(0)} + \frac{1}{v}q^{(-1)} + \ldots\ldots\ldots \qquad (4.4)$$

where $v^2 = \boldsymbol{z},$ \hfill (4.5)

and $\boldsymbol{z}$ is given by Eq. (3.5).

For the a very large value of $v$, from Eq. (4.4) it can be easily seen that

$$q^2 \approx v^2(q^{(1)})^2, \quad q^3 \approx v^3(q^{(1)})^3, \qquad (4.6)$$

and so on.

Substituting Eqs. (4.4) and (4.6) in Eq. (4.2) and ignoring the terms involving lower powers and retaining the terms involving the highest powers of $v$ we obtain

$$v^4 (q^{(1)})^2 \{(q^{(1)})^2 - a\} = 0. \qquad (4.7)$$

Solving Eq. (4.7), we get four values of $q^{(1)}$, given by

$$q_1^{(1)} = \sqrt{a}, \quad \overline{q}_1^{(1)} = -\sqrt{a}, \quad \text{and} \quad q_2^{(1)} = \overline{q}_2^{(1)} = 0, \qquad (4.8)$$

where $a > 0$, and is given by first equation in Eq. (4.3), and Eq. (3.6). In addition we see that in Eq. (4.8) $\overline{q}_1^{(1)}$ is not appearing as a complex conjugate of $q_1^{(1)}$, but bearing in mind Eq. (4.1) it could be adjusted where required.

In order to analyze the dispersion equation asymptotically, we substitute Eq. (4.4) in Eq. (3.9).

Avoiding the mathematical details, selecting the non-vanishing terms involving the highest powers of $v$, the expressions for
$f_1 \overline{g}_1$ and $\overline{f}_1 g_1$ are given by

$$f_1 \overline{g}_1 \approx i(a)^{-2} \overline{q}_1^{(1)} q_1^{(1)} [av^2 \{\overline{q}_1^{(1)}(\mathbf{1} + \mathbf{e}) + q_1^{(1)}(\mathbf{e} - \mathbf{1}^{-1})\} + \overline{q}_1^{(1)} q_1^{(1)} \{q_1^{(1)}(\mathbf{1} + \mathbf{e}) + \overline{q}_1^{(1)}(\mathbf{e} - \mathbf{1}^{-1})\}], \qquad (4.9)$$

$$\overline{f}_1 g_1 \approx i(a)^{-2} \overline{q}_1^{(1)} q_1^{(1)} [av^2 \{q_1^{(1)}(\mathbf{1} + \mathbf{e}) + \overline{q}_1^{(1)}(\mathbf{e} - \mathbf{1}^{-1})\} + \overline{q}_1^{(1)} q_1^{(1)} \{\overline{q}_1^{(1)}(\mathbf{1} + \mathbf{e}) + q_1^{(1)}(\mathbf{e} - \mathbf{1}^{-1})\}]. \qquad (4.10)$$

Subtracting Eq. (4.10) from Eq. (4.9), we get the asymptotic approximation of the first term in Eq. (3.9), given by

$$f_1\bar{g}_1 - \bar{f}_1 g_1 \approx ia^{-2}[av^2 - \bar{q}_1^{(1)} q_1^{(1)}](\bar{q}_1^{(1)} - q_1^{(1)})\bar{q}_1^{(1)} q_1^{(1)} \sqrt{e^2 + 4}. \tag{4.11}$$

Similarly we can obtain the approximations for other terms in Eq. (3.9) as

$$f_2\bar{g}_2 - \bar{f}_2 g_2 \approx ia^{-2}[av^2 - \bar{q}_2^{(1)} q_2^{(1)}](\bar{q}_2^{(1)} - q_2^{(1)})\bar{q}_2^{(1)} q_2^{(1)} \sqrt{e^2 + 4}, \tag{4.12}$$

$$f_1\bar{g}_2 - \bar{f}_2 g_1 \approx ia^{-2}[av^2 - \bar{q}_2^{(1)} q_1^{(1)}](\bar{q}_2^{(1)} - q_1^{(1)})\bar{q}_2^{(1)} q_1^{(1)} \sqrt{e^2 + 4}, \tag{4.13}$$

$$\bar{f}_1 g_2 - f_2 \bar{g}_1 \approx ia^{-2}[av^2 - q_2^{(1)} \bar{q}_1^{(1)}](q_2^{(1)} - \bar{q}_1^{(1)}) q_2^{(1)} \bar{q}_1^{(1)} \sqrt{e^2 + 4}, \tag{4.14}$$

$$\bar{a}_1 a_2 + a_1 \bar{a}_2 - 1 - a_1\bar{a}_1 a_2 \bar{a}_2$$
$$\approx 2\cosh[vkh(q_2^{(1)} - \bar{q}_1^{(1)})] - e^{vkh(\bar{q}_1^{(1)} - q_1^{(1)} + \bar{q}_2^{(1)} - q_2^{(1)})} - 1, \tag{4.15}$$

$$\bar{a}_1 a_1 + a_2 \bar{a}_2 - 1 - a_1 \bar{a}_1 a_2 \bar{a}_2$$
$$\approx [1 - e^{vkh(\bar{q}_2^{(1)} - q_2^{(1)})}][e^{vkh(\bar{q}_1^{(1)} - q_1^{(1)})} - 1], \tag{4.16}$$

where $q_1^{(1)}, \bar{q}_1^{(1)}, q_2^{(1)}, \bar{q}_2^{(1)}$ are given by Eq. (4.8).

Using Eqs. (4.11), (4.12), (4.13), (4.14), (4.15) and (4.16) in Eq. (3.9), the asymptotic approximation of the dispersion equation is given by

$$(a - \bar{q}_1^{(1)} q_1^{(1)})(a - \bar{q}_2^{(1)} q_2^{(1)})(\bar{q}_1^{(1)} - q_1^{(1)})(\bar{q}_2^{(1)} - q_2^{(1)})[2\cosh\{vkh(q_2^{(1)} - \bar{q}_1^{(1)})\} - 1$$

$$- e^{vkh(\bar{q}_1^{(1)} - q_1^{(1)} + \bar{q}_2^{(1)} - q_2^{(1)})}] + (a - \bar{q}_2^{(1)} q_1^{(1)})(a - q_2^{(1)} \bar{q}_1^{(1)})(\bar{q}_2^{(1)} - q_1^{(1)})(q_2^{(1)} - \bar{q}_1^{(1)})[1 -$$

$$e^{vkh(\bar{q}_2^{(1)} - q_2^{(1)})}][e^{vkh(\bar{q}_1^{(1)} - q_1^{(1)})} - 1] \approx 0. \qquad (4.17)$$

Substituting the values of $q_1^{(1)}$, $\bar{q}_1^{(1)}$, $q_2^{(1)}$, and $\bar{q}_2^{(1)}$ from Eq. (4.8) into Eq. (4.17) we then obtain

$$1 - e^{-2khv\sqrt{a}} \approx 0. \qquad (4.18)$$

*Keeping in view Eq. (4.1)*, replacing $\sqrt{a}$ by $i\sqrt{a}$ in Eq. (4.18) we obtain two sets of wave speeds as a function of wave number, layer thickness and material parameter as

$$v_n \approx \frac{n\boldsymbol{p}}{kh\sqrt{a}}, \qquad (4.19)$$

and

$$v'_n \approx \frac{n\boldsymbol{p}}{2kh\sqrt{a}}, \qquad (4.20)$$

where $n = 1,2,3,4,5,\ldots\ldots$

In Section 5 numerical results will be presented to illustrate the dependence of $v_n$ and $v'_n$ on the parameters $a$ and $kh$.

## 5. Numerical results

### 5.1 *CASE A:* $2\boldsymbol{b} = \boldsymbol{a} + \boldsymbol{g}$

By restricting the attention to a certain class of strain-energy functions namely those for which $2\boldsymbol{b} = \boldsymbol{a} + \boldsymbol{g}$, or, equivalently, $a = 1$, Eqs. (4.19) and (4.20) simplify to

$$v_n \approx \frac{n\boldsymbol{p}}{kh}, \tag{5.1}$$

$$v'_n \approx \frac{n\boldsymbol{p}}{2kh}, \tag{5.2}$$

where $n = 1,2,3,4,5,\ldots\ldots$.

Eqs. (5.1) and (5.2) are independent of the amount of shear and this is a different phenomenon in comparison with [6]. Fig.1 below shows the dependence of the velocities $v_n$ and $v'_n$ upon the wave number $k$ and layer thickness $h$ for different modes. All the figures have been produced using Mathematica [10].

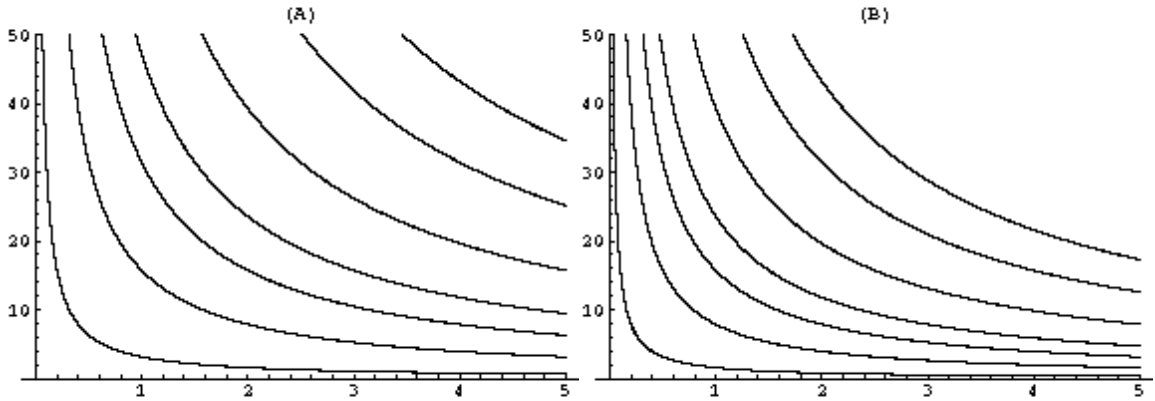

**Fig. 1.** For $n = 1,5,10,15,25,40,$ and $55,$

    (A) Plot of $v_n$ (vertical scale) against $kh$ (horizontal scale),

    (B) Plot of $v'_n$ (vertical scale) against $kh$ (horizontal scale).

## 5.2 CASE B: $2b \neq a + g$

In this case we illustrate the possible differences from Case A by considering a strain-energy function for which

$$b = \sqrt{ag}, \qquad (5.3)$$

which was employed in Ref.[11]. This equality is satisfied, for example, the Varga strain-energy function defined by

$$W = 2m(l_1 + l_2 + l_3 - 3), \qquad (5.4)$$

although this latter fact is not used explicitly. In (5.4) $m(>0)$ is the shear modulus.

Using (3.6), recalling first Eq. In (4.3), then on noting that $\dfrac{a}{g} = l^4$ in case of incompressible isotropic material, (4.19) and (4.20) take the form

$$v_n \approx \frac{2np}{kh\sqrt{e^2 + 4}}, \qquad (5.5)$$

$$v_n' \approx \frac{np}{kh\sqrt{e^2 + 4}}, \qquad (5.6)$$

respectively, where $e$ is given by (2.14) and $n = 1,2,3,4,5,\ldots\ldots$.

(5.5) and (5.6) show that the velocities do depend upon the amount of shear.

For different modes, the effect of simple shear, on the behaviour of the velocities $v_n$ and $v_n'$ is shown graphically in Fig. 2 and Fig. 3 respectively.

As we increase the amount of shear, velocity curves shrink. In addition do notice that the vertical and horizontal scales are fixed in Fig. 2 and Fig.3.

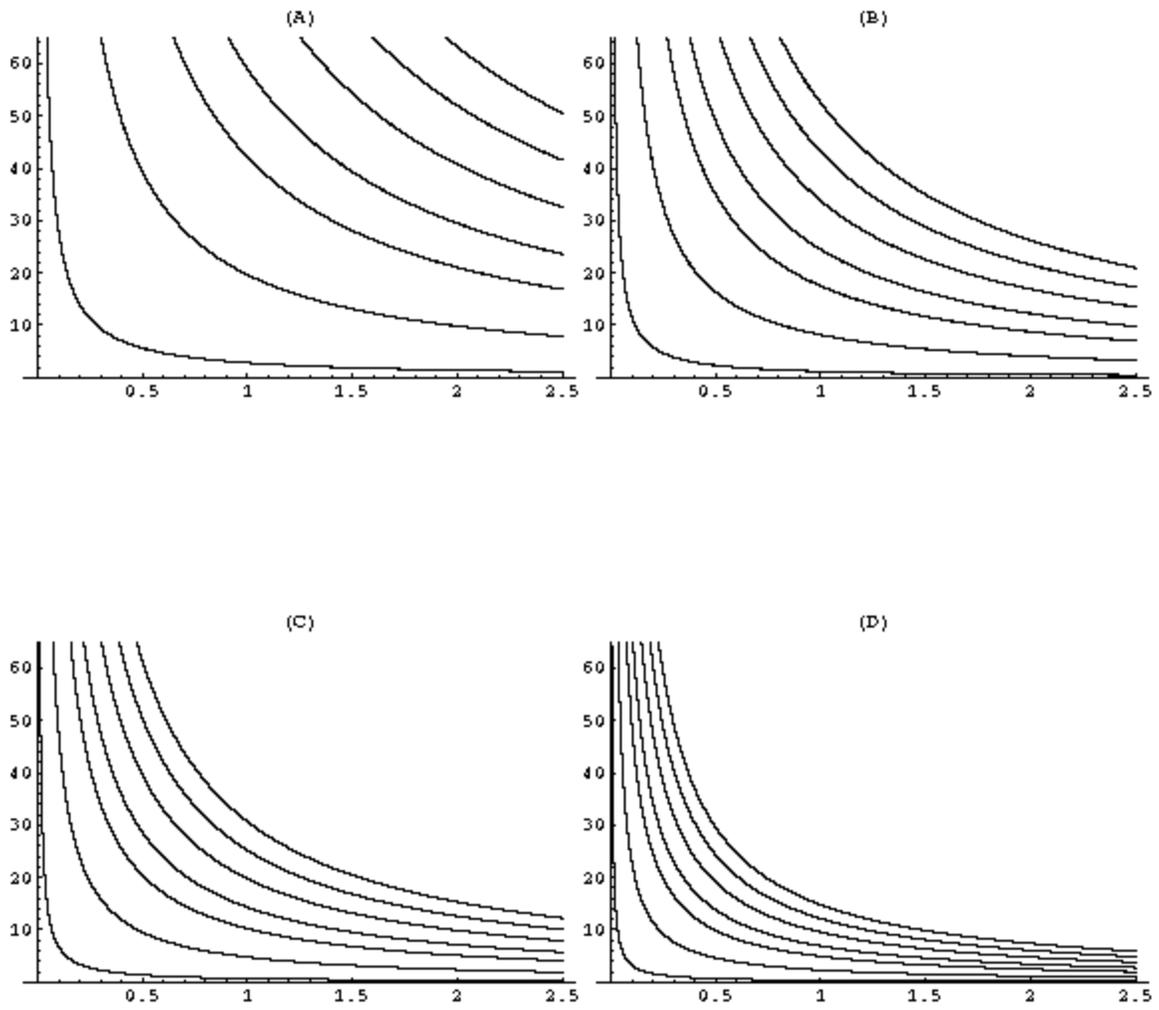

**Fig. 2.** Plot of $v_n$ (vertical scale) against $kh$ (horizontal scale) for $n = 1, 7, 15, 21, 29, 37, 45$: in

**Case B**: (A) $e = 1$; (B) $e = 5$; (C) $e = 9$; (D) $e = 19$.

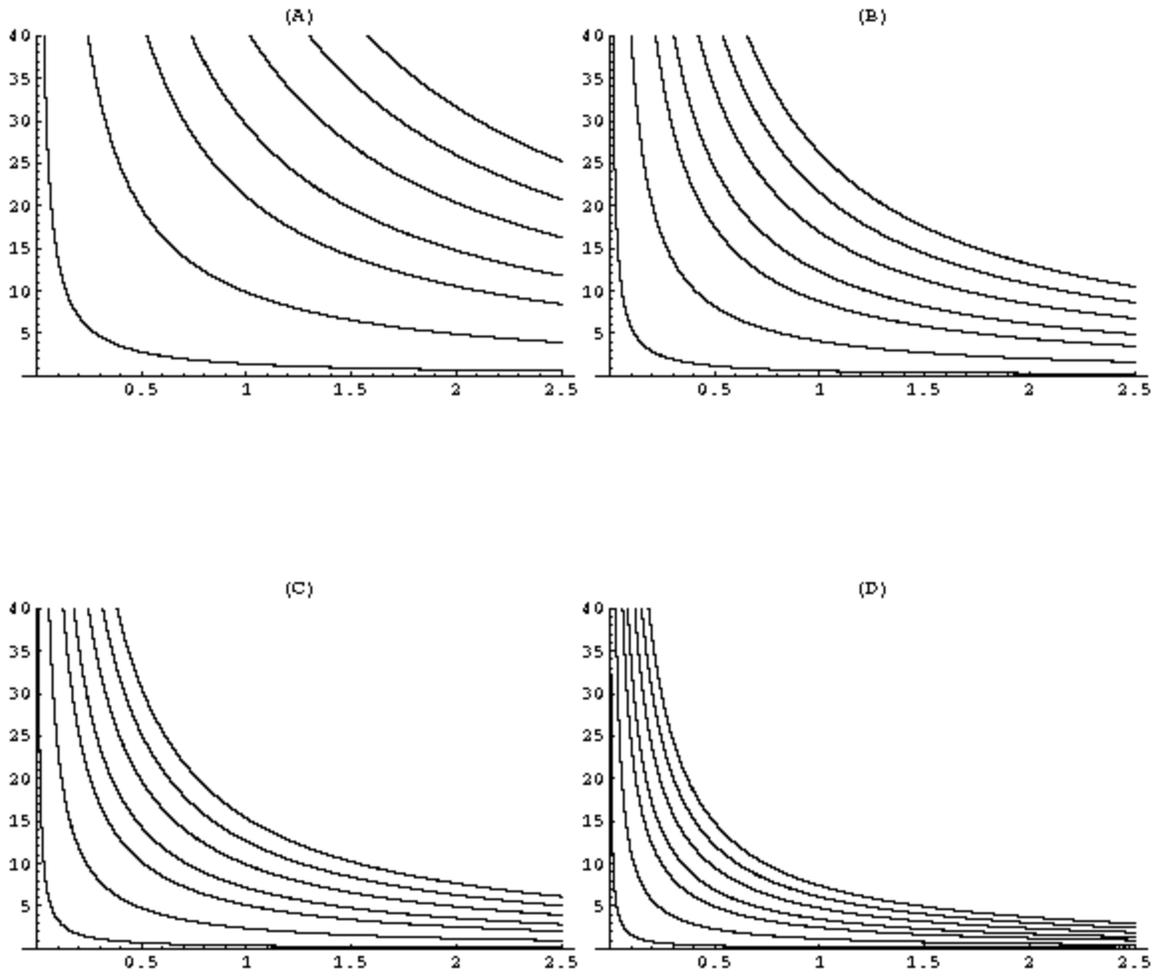

**Fig. 3.** Plot of $v'_n$ (vertical scale) against $kh$ (horizontal scale) for $n = 1, 7, 15, 21, 29, 37, 45$: in

Case **B**: (A) $e = 1$; (B) $e = 5$; (C) $e = 9$; (D) $e = 19$.